\documentclass{PoS}

\newcommand{\psib}{{\overline{\psi}}}

\title{Fermion Bags and A New Origin for a Fermion Mass}

\ShortTitle{New Origin for a Fermion Mass}

\author{\speaker{Shailesh Chandrasekharan}\thanks{Work done in collaboration with Venkitesh Ayyar and supported by US DoE grant DE-FG02-05ER41368.}\\
Physics Department, Duke University Durham NC 27704 \\
E-mail: \email{sch@phy.duke.edu}}

\abstract{The fermion bag is a powerful idea that helps to solve fermion lattice field theories using Monte Carlo methods. Some sign problems that had remained unsolved earlier can be solved within this framework. In this work we argue that the fermion bag also gives insight into a new mechanism of fermion mass generation, especially at strong couplings where fermion masses are related to the fermion bag size. On the other hand, chiral condensates arise due to zero modes in the Dirac operator within a fermion bag. Although in traditional four-fermion models the two quantities seem to be related, we show that they can be decoupled. While fermion bags become small at strong couplings, the ability of zero modes of the Dirac operator within fermion bags to produce a chiral condensate, can be suppressed by the presence of additional zero modes from other fermions. Thus, fermions can become massive even without a chiral condensate. This new mechanism of mass generation was discovered long ago in lattice field theory, but has remained unappreciated. Recent work suggests that it may be of interest even in continuum quantum field theory.}

\FullConference{The 32nd International Symposium on Lattice Field Theory,\\
23-28 June, 2014\\
Columbia University New York, NY}

\begin{document}

\section{Introduction}

One of the traditional paradigms of fermion mass generation in quantum field theory is spontaneous symmetry breaking. Perturbatively, fermion masses arise from local fermion bilinear terms in the action. In theories where symmetries forbid such terms, fermions can still become massive through spontaneous breaking of such symmetries. The symmetry breaking then manifests itself through a fermion bilinear condensate which we denote generically as $\langle \psib\psi\rangle \neq 0$. Can fermions become massive if some symmetry, which forbids fermion bilinear expectation values, remains unbroken? 

Anomaly matching severely constrains the chiral symmetries that can be preserved when fermions become massive \cite{tHooft}. The full chiral symmetry group of free fermions needs to be broken either explicitly or spontaneously for fermions to become massive. However, chiral symmetry subgroups can remain unbroken, which forbid fermion bilinear condensates yet allow fermions to become massive. Such exotic mechanisms of fermion mass generation have appeared in the literature in the context of QCD like theories \cite{Stern,Kogan} and more recently in condensed matter physics \cite{Slagle14}. In these examples the breaking of chiral symmetry that allows fermion to become massive, occurs through the appearance of four-fermion condensates.

Consider four-fermion field theories where interactions naturally generate the necessary four-fermion condensates that allow fermions to become massive, but still contain symmetries which forbid fermion bilinear condensates. In such theories there is no need for any further symmetry breaking for fermions to become massive. However, since four-fermion interactions are irrelevant perturbatively in three or more dimensions, there will still be a massless fermion phase at weak couplings. But at strong couplings, there can be a phase transition to a phase where fermions become massive without any spontaneous symmetry breaking. The strong coupling phase resembles the well known massive parity doubled phase \cite{Glozman}. Strictly speaking there is no local symmetry order parameter that distinguishes the two phases, although the four-fermion condensate is dramatically different in the two phases. Recently it has been argued that topological order plays an important role in distinguishing the two phases \cite{Slagle14}.

The exotic massive fermion phase is realized in a simple four-fermion field theory with two flavors of staggered fermions within the so called strong paramagnetic or PMS phase \cite{Hasenfratz,Lee}. In this work we show that the fermion bag approach sheds light on the mechanism through which such a phase is realized. We contrast two models both of which become massive at strong couplings. However, one allows the formation of a non-zero chiral condensate while the other forbids it.

\section{Model and Symmetries} \label{sec2}

The two models we study are simple four-fermion models constructed with staggered fermions. The first model (Model 1) is made with a single flavor of staggered fermions but with a nearest neighbor interaction whose Euclidean action is given by
\begin{equation}
S = \sum_{x,y} {\psib}_{x} \ M_{x,y} \ \psi_{y} - U \ \sum_{\langle xy\rangle} \Big\{ \psib_x\psi_x\ \psib_y\psi_y\Big\} 
\label{act1} 
\end{equation}
where $M_{x,y}$ is the free staggered fermion matrix. In addition to the usual space-time symmetries, the action is invariant under an $SU(2)\times U(1)$ subgroup of the full continuum chiral symmetry group. Under this subgroup the staggered fermion fields on even sites and odd sites transform as 
\begin{equation}
\left(\begin{array}{c} \psi_x \cr \cr \psib_x \end{array} \right) \rightarrow V \ \mathrm{e}^{i\theta}\  \left(\begin{array}{c} \psi_x \cr \cr \psib_x \end{array} \right)\ \ \mbox{and}\ \  
\left(\begin{array}{c} \psib_x \cr \cr  \psi_x \end{array} \right) \rightarrow V^* \ \mathrm{e}^{-i\theta}\  \left(\begin{array}{c} \psib_x \cr \cr \psi_x \end{array} \right),
\end{equation}
respectively. Note that the on-site chiral condensate $\psib_x\psi_x$ in not invariant under the $U(1)$ symmetry and hence cannot acquire a non-zero expectation value unless the symmetry is spontaneously broken. 

The second model (Model 2) is made with two flavors of staggered fermions with an onsite interaction whose Euclidean action is given by
\begin{equation}
S \ =\ \sum_{x,y}\ \Big\{ 
\overline{\psi^1_x} \ M_{x,y} \ \psi^1_y \ + 
\overline{\psi^2_x} \ M_{x,y} \ \psi^2_y \Big\}\  -\ U\ \sum_x \ \overline{\psi^1_x} \psi^1_x \ \overline{\psi^2_x} \psi^2_x.
\label{act2}
\end{equation}
This action is also invariant under the usual space-time symmetries. In addition it is also invariant under an $SU(4)$ subgroup of the continuum chiral symmetry group. Under this subgroup, the staggered fermion fields at even sites and odd sites transform as
\begin{equation}
\left(\begin{array}{c} \psi_{x,1} \cr \cr \overline{\psi_{x,1}} \cr \cr \psi_{x,2}  \cr \cr \overline{\psi_{x,2}}  \end{array} \right) \rightarrow V
\left(\begin{array}{c} \psi^1_x \cr \cr \overline{\psi^1_x} \cr \cr \psi^2_x  \cr \cr \overline{\psi^2_x}  \end{array} \right)\ \ \mbox{and} \ \  
\left(\begin{array}{c} \overline{\psi^1_x} \cr \cr  \psi^1_x \cr \cr \overline{\psi^2_x} \cr \cr \psi^2_x  \end{array} \right) \rightarrow V^*
\left(\begin{array}{c} \overline{\psi^1_x} \cr \cr  \psi^1_x \cr \cr \overline{\psi^2_x} \cr \cr \psi^2_x  \end{array} \right),
\end{equation}
respectively. In this model, the six onsite fermion bilinear condensates, $\psib_{x,1}\psi_{x,1}$, $\psib_{x,2}\psi_{x,2}$, $\psib_{x,1}\psi_{x,2}$, $\psib_{x,2}\psi_{x,1}$, $\psi_{x,1}\psi_{x,2}$, $\psib_{x,2}\psib_{x,1}$, transform as a sextet of $SU(4)$ and hence again are forbidden to acquire a non-zero expectation value unless the $SU(4)$ symmetry is spontaneously broken.

\section{The Fermion Bag Approach}

Fermion bags provide a new approach to write partition functions of lattice fermion field theories. In particular, there is no need to introduce auxiliary fields in four-fermion models. The weak coupling diagrammatic Monte Carlo methods for fermion systems developed recently, can be viewed as a subset of the fermion bag approach. A recent review can be found in \cite{Cha13}. Although the fermion bag approach was introduced as a new way to perform Monte Carlo calculations, here we show that it also gives new theoretical insight into the mechanism of fermion mass generation in the two models discussed above. Using fermion bags we will argue that at sufficiently large couplings, a chiral condensate can form in Model 1 but not in Model 2, although fermions in both models become massive.

\begin{figure}[t]
\vskip-0.5in
\begin{center}
\hbox{
\includegraphics[width=0.5\textwidth]{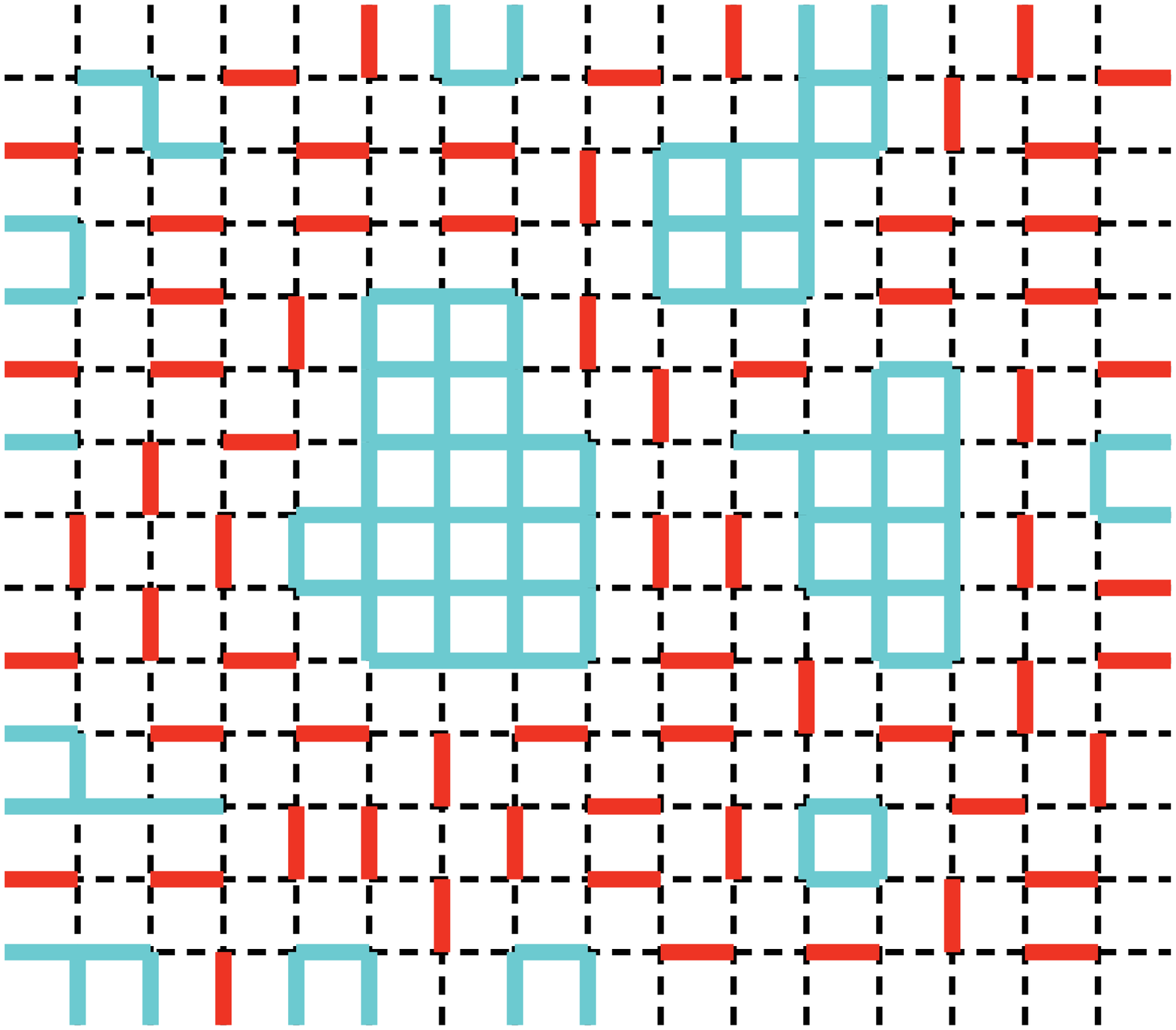}
\includegraphics[width=0.5\textwidth]{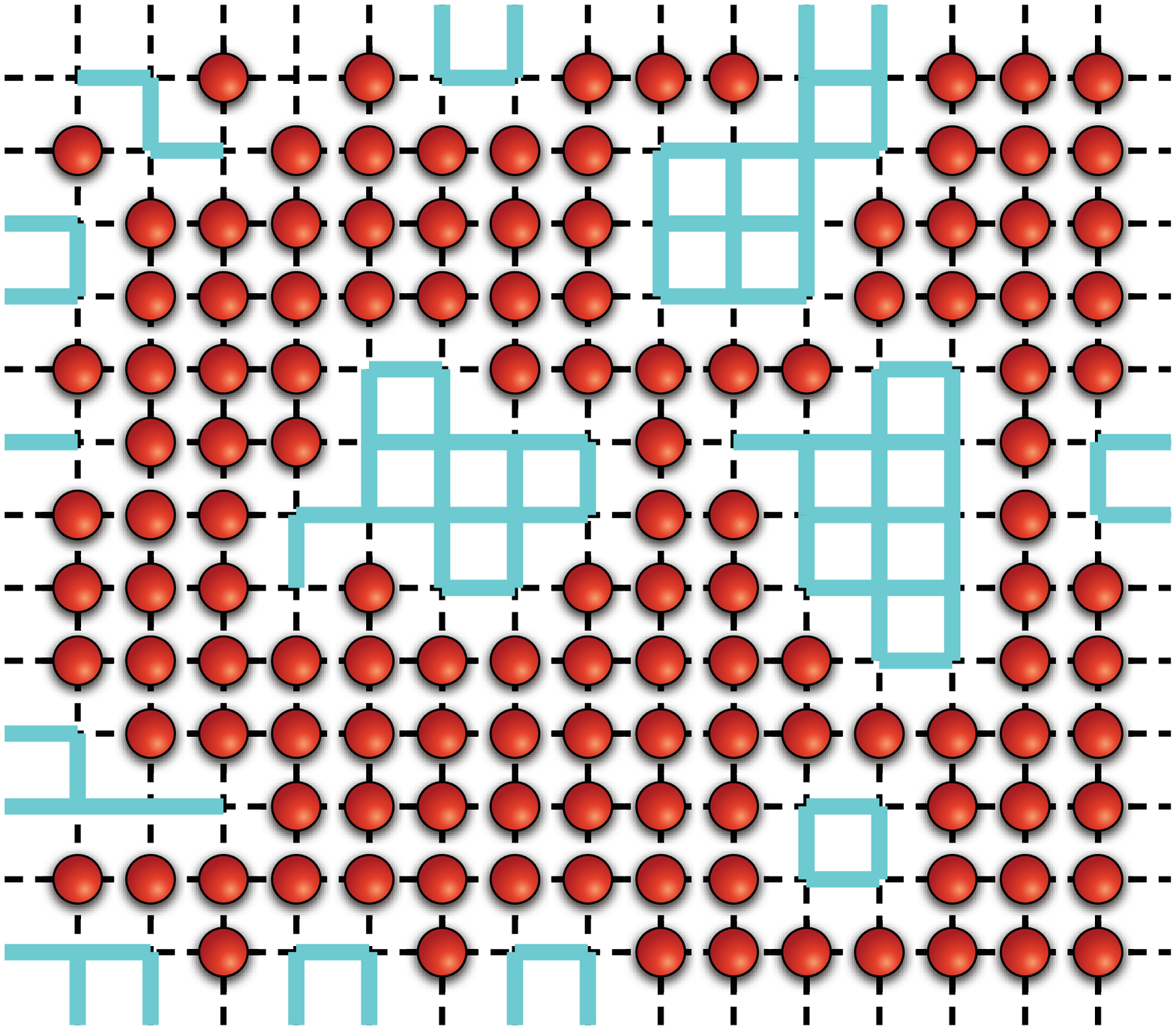}
}
\end{center}
\vskip-0.9in
\caption{\label{conf} Examples of fermion bag configurations in Model 1 (left) and Model 2 (right). Fermion bags are defined as the connected set of sites which do not belong to either a bond (left) or a monomer (right). $W_{\cal B}$ is the Dirac matrix associated with the bag.}
\end{figure}

In the fermion bag approach, the partition function of Model 1 can be written as
\begin{equation}
Z \ =\ \sum_{[b]} U^{N_B} \prod_{\cal B} \mathrm{Det}(W_{\cal B})
\end{equation}
where $[b]$ is a bond configuration (see Fig.\ref{conf} on the left) and $N_B$ is the number of bonds. Fermions can hop freely on sites that do not contain bonds. A connected region of such free sites forms a fermion bag denoted by ${\cal B}$. Fermion bags are determined uniquely in a given configuration $[b]$ and there can be many fermion bags. The matrix $W_{\cal B}$ is the Dirac matrix associated with the bag ${\cal B}$ and is obtained from the staggered fermion matrix $M$ by restricting it to sites within the bag. Similarly, the partition function of Model 2 can be written as
\begin{equation}
Z \ =\ \sum_{[n]} U^{N_m} \prod_{\cal B} \big\{\mathrm{Det}(W_{\cal B})\big\}^2
\label{mod2pf}
\end{equation}
where $[n]$ is a monomer configuration (see Fig.~\ref{conf} on the right) and $N_m$ is the number of monomers. The definition of fermion bags ${\cal B}$ and matrix $W_{\cal B}$ is the same as above, except that monomers replace bonds. The presence of two flavors is the reason for the square of the determinant in (\ref{mod2pf}).

\begin{figure}[t]
\begin{center}
\hbox{
\begin{minipage}[c]{0.5\textwidth}
\includegraphics[width=\textwidth]{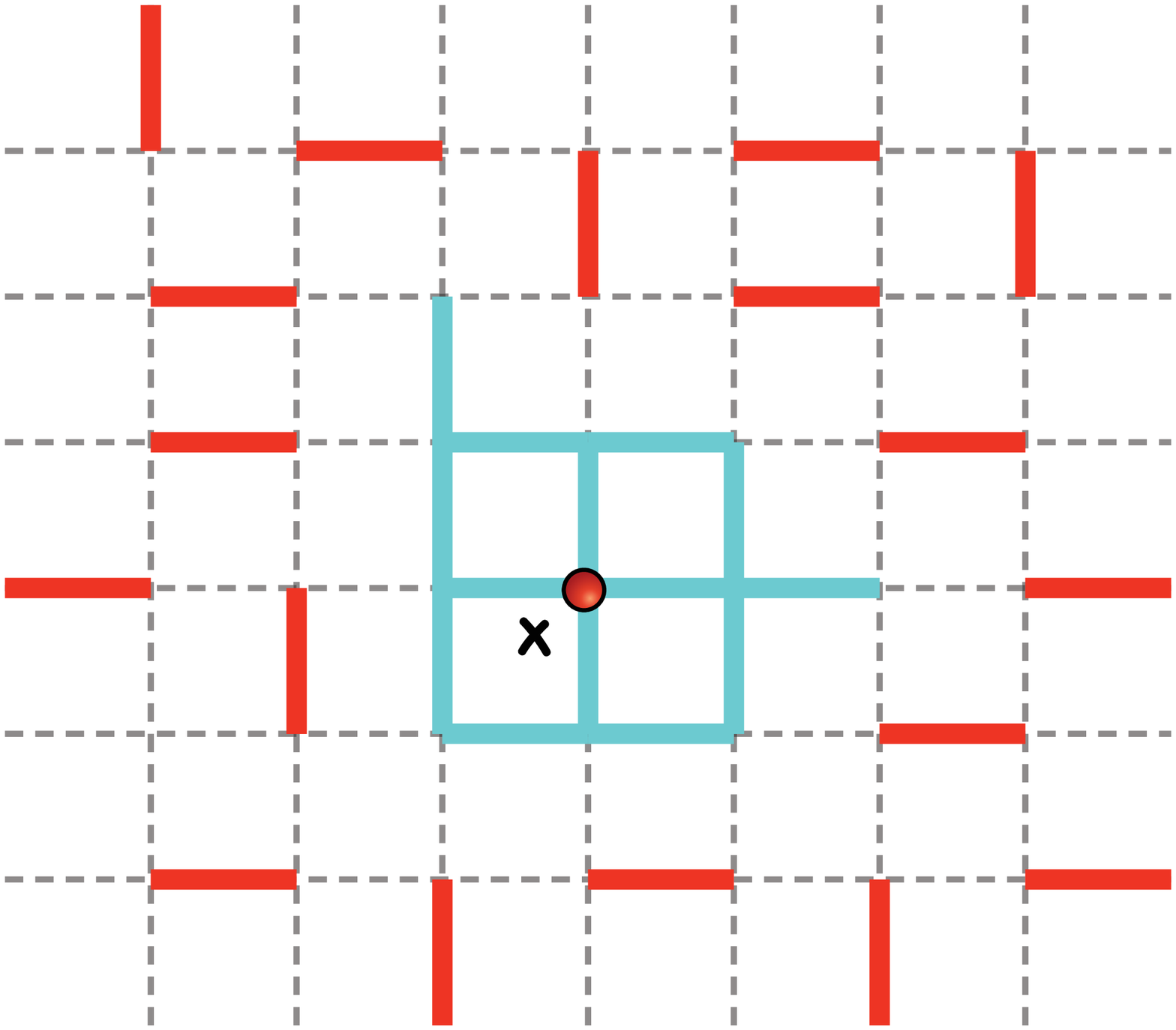}
\end{minipage}
\begin{minipage}[c]{0.49\textwidth}
\includegraphics[width=\textwidth]{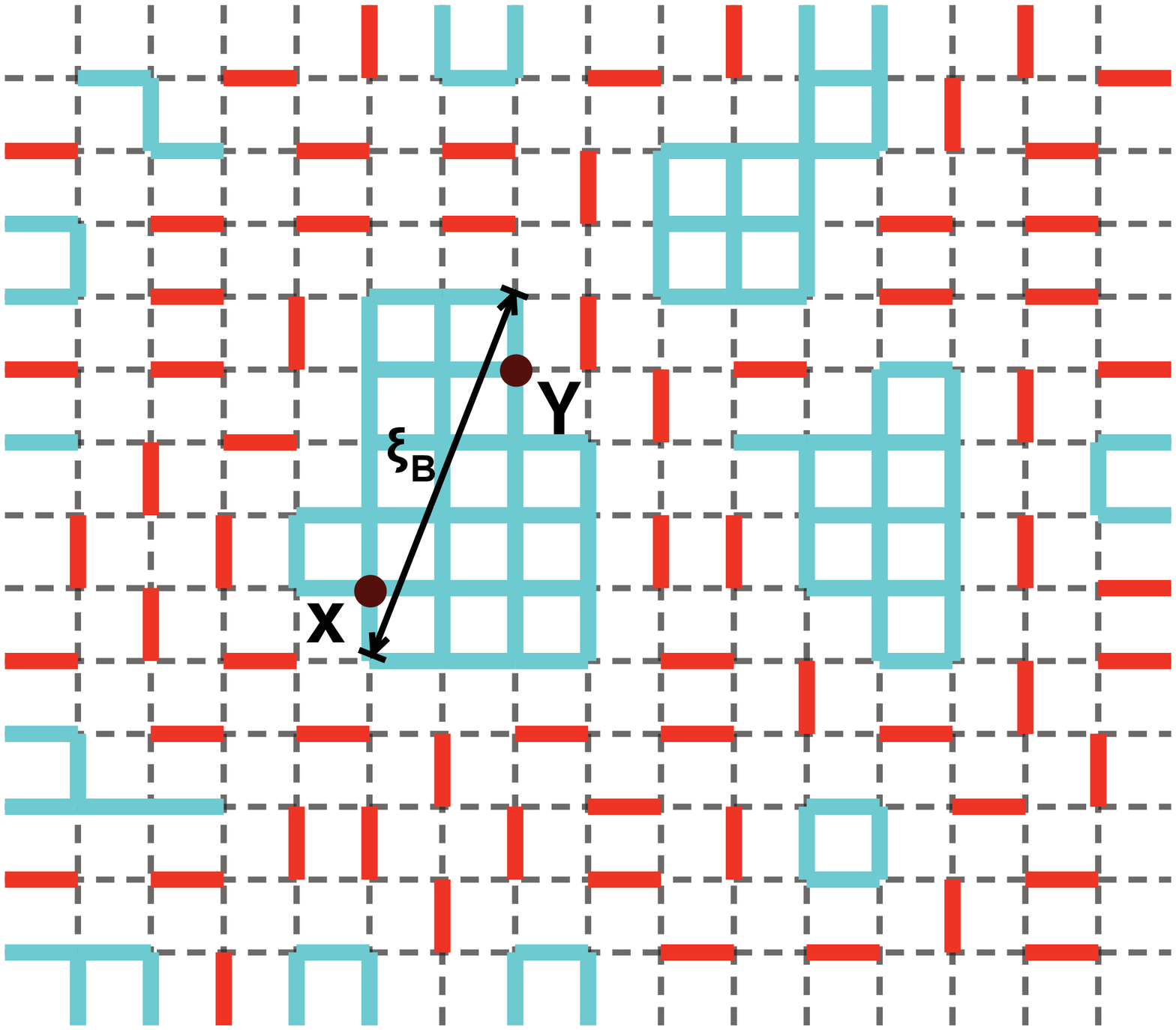}
\end{minipage}
}
\hbox{
\begin{minipage}[c]{0.48\textwidth}
\caption{\label{zmode} An example of a $\nu=1$ topological fermion bag. Since it contains the site $x$ it is labelled as ${{\cal B}_x}$. $W_{{\cal B}_x}$ will have one zero mode. Such topological bags contribute to the chiral condensate in Model 1, but not in Model 2.}
\end{minipage}
\hskip0.15in
\begin{minipage}[c]{0.48\textwidth}
\caption{\label{fmass}. Fermion two point correlation functions get contribution from points within the fermion bag. $\xi_{\cal B}$ is defined as the largest distance between two points in the fermion bag.}
\end{minipage}
}
\end{center}
\end{figure}

An important property of $W_{\cal B}$ is that it is anti-symmetric with non-zero matrix elements only between even and odd sites of the bag. Hence, in a bag which does not contain an equal number of even and odd sites, $W_{\cal B}$ will have zero modes. We can introduce the concept of ``topology'' of a bag by defining its topological charge as $\nu = n_e - n_o$, where $n_e$ ($n_o$) is the number of even (odd) sites in the bag.  Then $W_{\cal B}$ will have $|\nu|$ zero modes for topological reasons, very similar to the index theorem in the continuum. An example of a $\nu = 1$ topological fermion bag is shown in Fig.~\ref{zmode}.

\section{Fermion Mass}
 
Fermion masses are computed through the exponential decay of fermion two point correlation functions, defined through the relation
\begin{equation}
\Big\langle \psi_{x,1} \ \psib_{y,1} \Big\rangle \ =\ \frac{1}{Z}\ \sum_{[n]}  \ G_{{\cal B}'} (x,y) \ U^{N_m}\ \prod_{\cal B} \big\{\mathrm{Det}(W_{\cal B})\big\}^2,
\end{equation}
in Model 2. Here the sum is only over those configurations $[n]$ where both $x$ and $y$ belong to the same bag ${\cal B}'$ and the fermion propagator $G_{{\cal B}'}(x,y)$ is obtained by computing the appropriate matrix element of $(W_{{\cal B}'})^{-1}$. A similar relation exists in Model 1.

If $\xi_{{\cal B}}$ is defined as the largest possible distance between any two points within a bag (see Fig.~\ref{fmass}) then the fermion mass (in lattice units) must be greater than or equal to the ``typical'' values of $1/\xi_{\cal B}$ in fermion bags sampled during the Monte Carlo calculation. In other words, the ``typical'' fermion bag size determines the fermion mass. At strong couplings when fermion bags are small, fermions become massive.

\section{Chiral Condensate}

Traditionally we expect a non-zero fermion mass to exist only in the presence of a chiral condensate, i.e., $\langle \psib\psi\rangle \neq 0$. As we will see below, this is not necessary. The chiral condensate can be computed using the large distance behavior of the two point bosonic correlation function through the relation
\begin{equation}
\lim_{|x - y| \rightarrow \infty} \ \langle \overline{\psi_x}\psi_x \ \overline{\psi_y}\psi_y\rangle =  \langle \overline{\psi}\psi \rangle^2.
\label{cc}
\end{equation}
In the fermion bag approach the bosonic correlation function can get contributions from a connected and a disconnected component. In the connected component both $x$ and $y$ belong to the same bag, while in the disconnected component $x$ and $y$ belong to different bags. In Model 1 these two components are given by
\begin{eqnarray}
\Big\langle \overline{\psi_x} \psi_x \ \overline{\psi_y} \psi_y \Big\rangle_{\rm Conn} \ &=& \ \frac{1}{Z}\ \sum_{[b]} \ \Bigg\{  \ U^{N_B}\ \mathrm{Det}(W_{{\cal B}_{x,y}}([x,y])) \ \prod_{ {\cal B} \neq {\cal B}_{x,y}} \big\{\mathrm{Det}(W_{\cal B})\big\} \ \Bigg\}
\\
\Big\langle \overline{\psi_x} \psi_x \ \overline{\psi_y} \psi_y \Big\rangle_{\rm Disc} \ &=& \ \frac{1}{Z}\ \sum_{[b]}\ \Bigg\{ \ U^{N_B}\mathrm{Det}(W_{{\cal B}_x}([x]))\ \mathrm{Det}(W_{{\cal B}_y}([y]))
\prod_{ {\cal B} \neq {\cal B}_x,{\cal B}_y} \big\{\mathrm{Det}(W_{\cal B})\big\}\Bigg\}.
\label{ccmod1}
\end{eqnarray}
In the above expressions ${\cal B}_{x,y}$ stands for the fermion bag that contains both the sites $x$ and $y$, ${\cal B}_x$ is the bag that contains $x$ but not $y$, similarly ${\cal B}_y$ is the bag that contains $y$ but not $x$. Further, $W_{{\cal B}_{x,y}}([x,y])$ is the fermion bag Dirac matrix with sites $x$ and $y$ removed, similarly $W_{{\cal B}_x}([x])$ is the fermion bag Dirac matrix with the site $x$ dropped and similarly $W_{{\cal B}_y}([y])$ is the matrix with the site $y$ dropped. The connected and disconnected components in Model 2 are given by
\begin{eqnarray}
\Big\langle \overline{\psi^1_x} \psi^1_x \ \overline{\psi^1_y} \psi^1_y \Big\rangle_{\rm Conn} &=&
\frac{1}{Z}\ \sum_{[n]} \ \Bigg\{ \ U^{N_m} \ \mathrm{Det}(W_{{\cal B}_{x,y}}([x,y]))\  \mathrm{Det}(W_{{\cal B}_{x,y}})
\prod_{ {\cal B} \neq {\cal B}_{x,y}} \big\{\mathrm{Det}(W_{\cal B})\big\}^2\Bigg\}
\\
\Big\langle \overline{\psi^1_x} \psi^1_x \ \overline{\psi^1_y} \psi^1_y \Big\rangle_{\rm Disc} &=& \frac{1}{Z}\ \sum_{[n]} \ \Bigg\{  \ U^{N_m}\ 
\mathrm{Det}(W_{{\cal B}_x}([x]))\ \mathrm{Det}(W_{{\cal B}_x})\ 
\mathrm{Det}(W_{{\cal B}_y}([y]))\ \mathrm{Det}(W_{{\cal B}_y})
\nonumber \\
&&\ \ \ \ \ \ \ \ \ \ \ \ \ \ \times \
 \prod_{ {\cal B} \neq {\cal B}_x,{\cal B}_y} \big\{\mathrm{Det}(W_{\cal B})\big\}^2\ \Bigg\}.
\label{ccmod2}
\end{eqnarray}
Since at strong couplings large bags are exponentially suppressed, the connected component of the two point correlation function (\ref{cc}) will decay exponentially. Hence, the condensate can only form if the disconnected component can contribute to the correlation function, especially when $x$ and $y$ are located in two different fermion bags far from each other. 

It is easy to argue that in Model 1, the disconnected component (\ref{ccmod1}) is non-zero due to the existence of $\nu=\pm 1$ fermion bags in which $\mathrm{Det}(W_{{\cal B}_x}([x])) \neq 0$ although $\mathrm{Det}(W_{{\cal B}_x}) = 0$ (see Fig.~\ref{zmode}). The feature of the zero mode that it contributes to the condensate, but not to the partition function is well known in the continuum \cite{Leut92}. In contrast, in Model 2 the right hand side of (\ref{ccmod2}) exactly vanishes since $\nu=\pm 1$ bags satisfy the identity $\mathrm{Det}(W_{{\cal B}_x}([x]))\ \mathrm{Det}(W_{{\cal B}_x}) = 0$. In other words, if the zero mode from one flavor tries to contribute to the condensate, the same zero mode present in the other flavor forbids it. Since the disconnected part cannot contribute, at sufficiently strong couplings the chiral condensate vanishes in Model 2 although fermions are massive.

\begin{figure}[t]
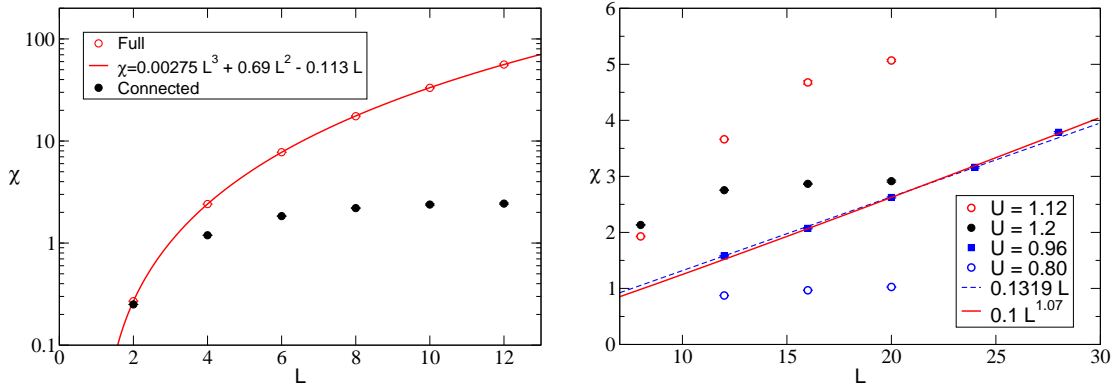

\begin{center}
\hbox{
\includegraphics[width=0.47\textwidth]{susmodel1.eps}
\hskip0.2in
\includegraphics[width=0.46\textwidth]{susmodel2.eps}
}
\caption{\label{results} Plots showing the susceptibility as a function of lattice size for Model 1 at $U = 0.8$ (left) and Model 2 at different values of $U$ (right). In Model 1 we distinguish between the full and the connected component of the susceptibility.}
\end{center}
\end{figure}

\section{Numerical Results}

The discussion presented above is valid at large $U$ To understand how large is large enough, we have computed the full susceptibility and its connected component defined as
\begin{equation}
\chi(L) = \frac{1}{L^3}\sum_{x,y} \Big\langle \overline{\psi_x} \psi_x\ \overline{\psi_y}\psi_y \Big\rangle,
\ \ \ \ \ \ \chi_{\rm Conn}(L) = 
\frac{1}{L^3}\sum_{x,y} \Big\langle \overline{\psi_x} \psi_x\ \overline{\psi_y}\psi_y \Big\rangle_{\rm Conn}.
\end{equation}
In the previous section we predicted that these two quantities behave differently at large $U$ in Model 1. While the full susceptibility should diverge with the volume, the connected component should saturate. In Fig.~\ref{results} we show that this is observed even at $U=0.8$. The transition to the massless phase occurs at $U_c=0.26$. Similarly, we predicted that in Model 2 that the full susceptibility will saturate at large couplings. Fig.~\ref{results} shows that this occurs even at $U=1.2$. In this case the transition to the massless phase occurs at $U=0.96$ where $\chi(L)$ diverges due to the presence of a second order transition. 

\section{Conclusions}

The fermion bag approach sheds light on a new mechanism of fermion mass generation. While the fermion mass is related to the fermion bag size, the chiral condensate arises due to special topological fermion bags in which the Dirac operator has zero modes. There are interesting lattice fermion models where typical fermion bags are small and topological bags are highly suppressed. In such models, fermions become massive without the formation of a chiral condensate. Recent results suggest that this exotic fermion mass generation mechanism could be of interest in continuum quantum field theory \cite{Ayyar}.

\end{document}